\documentclass[authorversion,nonacm]{acmart}
\usepackage{graphicx} 

\title{Establishing Trust in Crowdsourced Data}

\date{November 4 2025}

\setcopyright{none}

\author{Iffat Gheyas}
\affiliation{%
  \institution{Surrey Centre for Cyber Security, University of Surrey}
  \country{UK}}
\email{iagheyas@hotmail.co.uk}

\author{Muhammad Rizwan Asghar}
\affiliation{%
  \institution{Surrey Centre for Cyber Security, University of Surrey}
  \country{UK}}
\email{r.asghar@surrey.ac.uk}

\author{Steve Schneider}
\affiliation{%
  \institution{Surrey Centre for Cyber Security, University of Surrey}
  \country{UK}}
\email{s.schneider@surrey.ac.uk}

\author{Alan Woodward}
\affiliation{%
  \institution{Surrey Centre for Cyber Security, University of Surrey}
  \country{UK}}
\email{alan.woodward@surrey.ac.uk}

\begin{document}

\begin{abstract}
    Crowdsourced data enables real-time decision-making but faces several challenges, including misinformation, errors, and dominance by a few contributors. This study systematically examines trust management practices across platforms categorised as Volunteered Geographic Information, Wiki Ecosystems, Social Media, Mobile Crowdsensing, and Specialised Review and Environmental Crowdsourcing. Key strengths of current techniques and practices are automated moderation and community validation, while limitations involve rapid data influx, niche oversight gaps, opaque trust metrics, and elite dominance. We discuss proposed solutions that incorporate advanced AI tools, transparent reputation metrics, decentralised moderation, structured community engagement, and a ``soft power'' strategy, aiming to equitably distribute decision-making authority and enhance overall data reliability. 
\end{abstract}

\maketitle

\section{Introduction}
Crowdsourced data has rapidly emerged as a cornerstone for real-time decision-making, leveraging collective insights from diverse contributors worldwide to facilitate timely, detailed, and contextually rich information. This capability significantly enhances disaster response efficiency, inclusivity in policy-making, and resource allocation effectiveness \cite{item1}. Crowdsourced data refers to information collaboratively created, refined, or validated by distributed groups rather than a centralised authority. This decentralisation inherently reduces potential biases and mitigates delays common in traditional, hierarchical data collection processes. Contributors typically submit their observations via online platforms, such as collaborative mapping services -- e.g., OpenStreetMap (OSM) -- and social media networks, generating extensive pools of user-driven content that reflect real-time conditions and local perspectives. Motivations driving contributors are diverse, encompassing civic altruism, community recognition, social engagement, and the prospect of enhanced local services supported by accurate, timely data.

Real-world events highlight the immense value of crowdsourcing, especially in high-stakes scenarios, such as conflict zones and humanitarian emergencies. In the early stages of the war in Ukraine, individuals leveraged platforms such as Telegram and X (formerly Twitter) to document troop movements, infrastructure damage, and urgent civilian needs, significantly aiding military analysts and humanitarian agencies \cite{item2}, \cite{item3}. Similarly, during severe flooding in the UK, citizens used crisis-mapping tools—often built upon OSM or integrated into local volunteer apps—to report rising water levels and blocked roads, enabling emergency services to optimise rescue efforts \cite{item4}. Crowdsourced initiatives have also been instrumental in conflict-affected regions of the Middle East, where volunteers utilise platforms, such as the Humanitarian OpenStreetMap Team (HOT), to map safe evacuation routes and identify critical resources in near real-time \cite{item5}. By empowering local observers to provide timely, precise information, these crowdsourced tools equip decision-makers with actionable intelligence essential for protecting lives, efficiently allocating resources, and responding effectively to rapidly evolving crises.

Despite several benefits, the inherent openness of crowdsourced channels also leaves them susceptible to rapidly spreading misinformation, posing critical challenges for policymakers in the UK, the US, and Europe. Early in the COVID-19 pandemic, conspiracy theories linking 5G technology to the virus proliferated in UK-based Facebook groups, prompting arson attacks on telecommunications infrastructure and necessitating corrective public communications by authorities \cite{item6}. Similarly, in the US, the dissemination of the ``Stop the Steal'' narrative via X culminated in the January 6 Capitol riots, triggering significant law enforcement and National Guard mobilisation in response to unfounded election fraud claims \cite{item7}. Additionally, coordinated Telegram channels across Europe spread disinformation about the Ukraine conflict in 2022, compelling the European External Action Service to issue multiple clarifications and advocate reliance on verified information sources \cite{item8}. These examples underscore how platforms designed for enhanced situational awareness can be exploited to propagate deceptive content at scale, undermining public trust and complicating rapid, informed responses by governmental and defence entities.

While crowdsourced data offers transformative potential, its openness also heightens vulnerability to misinformation, making data trustworthiness critical. Trustworthiness refers to the degree to which crowd-contributed data can be considered accurate, reliable, and credible \cite{item9}, \cite{item10}. Decision-makers, from scientists and public administrators to business strategists, depend heavily on trustworthy data due to its far-reaching implications. Researchers commonly categorise trustworthiness into three interconnected dimensions: accuracy, reliability, and credibility \cite{item9}, \cite{item10}.
Accuracy reflects how closely crowdsourced information matches objective reality or ``ground truth'' \cite{item9}. Its definition varies across platforms. In Volunteer Geographic Information (VGI) projects, such as OSM, accuracy typically aligns with ISO 19157 measures: positional accuracy, thematic accuracy, completeness, logical consistency, and temporal accuracy \cite{item11}. Within the wiki ecosystem (e.g., Wikipedia, Wikidata), accuracy primarily involves verifiability through cited references and editorial processes such as ``talk page'' discussions and revision histories \cite{item12}. In social media, accuracy is often evaluated via platform moderators, third-party fact-checkers, or community-driven flags \cite{item13}. For crowdsensing applications collecting sensor data from personal devices, accuracy depends on calibration against official standards or aggregating multiple sensor inputs to identify anomalies and improve overall measurement quality \cite{item14}.

Reliability refers to the consistency and stability of crowdsourced information over time, emphasising repeated contributions and convergence among multiple independent observations \cite{item15}. In crowdsourced geographic data contexts, reliability specifically refers to the temporal stability of data and the degree of consensus among multiple edits or contributions regarding the same geographic instance. Provenance trails—documenting who contributed, when, and how—are crucial for assessing this aspect. For example, Wikipedia enhances reliability through its ``no original research'' policy, requiring corroboration by independent contributors \cite{item16}. Similarly, crowdsensing applications rely on repeated sensor readings from various devices to establish reliable data \cite{item17}. Thus, reliability, underpinned by temporal consistency and multi-source consensus, significantly contributes to the overall trustworthiness of crowdsourced data.

Credibility (or reputation) reflects the perceived trustworthiness of data contributors or sources \cite{item9}. Credibility increases with certain factors, such as subject-matter expertise, historical accuracy, and recognition by respected community members. Wikipedia contributors build credibility through constructive edits supported by authoritative references \cite{item18}. On social media platforms, verified accounts and recognised experts typically hold higher credibility, though verification alone is not fail-safe \cite{item19}. By jointly evaluating accuracy, reliability, and credibility, researchers and practitioners can better assess whether crowdsourced data provides sufficient confidence for critical decision-making, ensuring collective insights effectively translate into practical benefits.

Crowdsourcing platforms—ranging from collaborative mapping initiatives and user-generated wikis to social media, mobile sensing apps, and specialised reporting sites—share a critical need to manage data trustworthiness through accuracy, reliability, and contributor credibility. Central to this is the implementation of trust management systems \cite{item20}, tailored uniquely across platforms through distinct tools and policies. These systems assess the quality of each contribution based on its accuracy and reliability, adjusting contributors' reputation scores accordingly: high-quality submissions improve contributors' credibility, while flawed or misleading content diminishes it. This reputation directly influences the perceived trustworthiness and subsequent impact of contributors’ future submissions, forming a reinforcing feedback loop. Higher-rated contributors thus exert greater influence over final platform outputs, while lower-rated users find their submissions less influential and subject to increased scrutiny. This dynamic motivates even previously less trustworthy contributors to enhance their content quality over time, ultimately fostering platform-wide improvements in data reliability for informed and effective decision-making.

Establishing trust in crowdsourced data is inherently challenging, primarily because the ``ground truth'' is often unobservable in real time and may remain inaccessible for extended periods \cite{item21}. The dynamic and heterogeneous nature of crowdsourced data further complicates accuracy assessments, requiring careful consideration of numerous factors when evaluating both contribution quality and contributor intent \cite{item22}. This report synthesises the current state of practice in evaluating and enhancing trust across crowdsourcing platforms, examining how existing methods—both those directly employed by platforms and those discussed in the literature—address or fall short in tackling critical challenges such as misinformation, quality control, and contributor reputation.

In this work, we examine current trust management practices across six major categories of crowdsourcing platforms, highlighting how prominent platforms manage data quality and community oversight. To our knowledge, no existing review has comprehensively explored trust management across such diverse crowdsourcing categories. Platforms reviewed include: VGI represented by OSM (Section 3); Wiki ecosystems exemplified by Wikipedia and Wikidata (Section 4); social media platforms X and YouTube (Section 5); mobile crowd-sensing applications including Waze, OpenSignal, Sensorly, Safecast, and AirCasting (Section 6); specialised environmental crowdsourcing platforms such as MyCoast and Galaxy Zoo (Zooniverse) (Section 7); and specialised review crowdsourcing platforms like Yelp, TripAdvisor, and Amazon’s Customer Review System (Section 8). The study combines insights from official platform documentation, supplementary online resources, and a systematic literature review (detailed in Section 2, Methodology). Subsequent sections synthesise key strengths, limitations, and emerging solutions. Finally, Section 9 summarises our findings and provides concluding remarks, offering practical recommendations to enhance trust management systems across diverse crowdsourcing domains.

\section{Research Methodology}

This study employs a two-part methodology combining an analysis of current trust management practices by prominent crowdsourcing platforms and a systematic literature review to identify emerging solutions for improving these practices. The research aims to answer the following research questions:
\begin{itemize}
    \item 
What are the strengths of the current trust management practices employed by prominent crowdsourcing platforms?
\item What are the limitations and challenges of these current trust management practices?
\item What are the opportunities for future advancements in trust management systems for crowdsourcing platforms?
\end{itemize}
\subsection{Current State-of-practice Analysis}
To understand the current trust management practices employed by leading crowdsourcing platforms (e.g., OSM, Wikipedia, X, YouTube), we conducted an extensive review primarily using publicly available information from official platform websites, documentation, and supplementary online resources. Additionally, selected academic papers identified during preliminary literature searches were consulted to enrich and clarify our analysis, though these were not part of the systematic literature review. Targeted Google searches utilised specific keywords including ``trust management,'' ``content moderation,'' ``misinformation policies,'' ``community guidelines,'' ``user credibility,'' ``fake review detection,'' and ``data reliability.'' Information gathered from these diverse sources was systematically synthesised to provide a comprehensive overview of each platform's documented strategies, tools, and procedures, highlighting key strengths, limitations, and challenges.

\subsection{Systematic Literature Review for Emerging Solutions}
To identify and evaluate emerging solutions (policies and algorithms) proposed within academic literature, we performed a systematic literature review guided by the PRISMA (Preferred Reporting Items for Systematic Reviews and Meta-Analyses) protocol \cite{item23}. PRISMA provides clear guidelines for structuring systematic reviews, ensuring transparency, replicability, and thoroughness. The selection process comprised four sequential stages: Identification, Screening, Eligibility, and Inclusion.

\paragraph{Identification} 
We searched five scholarly databases: Google Scholar, ACM Digital Library, IEEE Xplore Digital Library, Scopus, and Web of Science. The following keywords guided our searches: ``VGI'', ``OpenStreetMap'', ``Wikipedia'', ``Wikidata'', ``Social Media'', ``YouTube'', ``X'', ``Twitter'', ``Mobile Crowdsensing Platforms,'' ``Review crowdsourcing'' combined with ``trustworthiness'', ``misinformation'', ``vandalism'', ``false data'', and ``error''. Initially, we retrieved up to the top 100 articles from each database for each search term (or all available articles if fewer than 100 were returned per search term), yielding a total of 896 records for subsequent screening.

\paragraph{Screening} 
The screening phase applied stringent exclusion criteria sequentially to shortlist relevant papers: 1) Removal of duplicate records; 2) Exclusion of non-English papers; 3) Exclusion of articles published before January 2016 (chosen to balance practical manageability and include recent, advanced solutions reflecting current trends and technologies); and 4) Exclusion of review or survey papers. Papers were retained if they met all the following inclusion criteria based on title and abstract review: 1) Published in peer-reviewed journals, conference proceedings, or book chapters; 2) Explicitly suggested policies or methodologies for promoting trustworthy content through misinformation detection and prevention; and 3) Full texts accessible via the University of Surrey library. After screening, 268 papers proceeded to the eligibility stage. Paper selection and review continued up to February 2025, capturing the most current literature available at the time of writing.

\paragraph{Eligibility} 
In the eligibility phase, we conducted a detailed review of each paper's full text to ensure relevance. Articles were included if they proposed innovative and practical solutions explicitly aimed at enhancing trust management systems within crowdsourcing platforms.

\paragraph{Inclusion} 
Ultimately, 55 papers offering actionable recommendations and innovative methods for improving crowdsourcing trust management were included in our final synthesis and review.

\section{VGI: State of Practice in Trust Management}

VGI, a term coined by Goodchild (2007), involves individuals --- often non-experts --- contributing local knowledge to create collective spatial datasets \cite{item24}. Enabled by Web 2.0 technologies, VGI lowers barriers to sharing geospatial data, such as points of interest, roads, and buildings. A key driver of projects like OSM is to provide freely accessible geographic data, countering the restrictive licenses and costs of traditional mapping agencies like the Ordnance Survey \cite{item25}. VGI enhances map coverage and detail, especially in underserved regions, through user participation and open access.

OSM, launched in 2004, is the largest VGI initiative, offering a free, editable world map \cite{item25}. Registered users edit geospatial features—nodes (points), ways (lines/polygons), and relations (complex links)—using tools like iD, Potlatch, and JOSM. Its global community, including local mappers and remote contributors digitising satellite imagery, continuously updates roads, buildings, and amenities. OSM supports diverse users, from government agencies to humanitarian groups, for navigation and disaster relief. As a leading open-data alternative to proprietary maps, OSM demonstrates the power of crowdsourcing to fill gaps in official datasets.

OSM’s data structure comprises nodes, ways, and relations, tagged with key-value pairs (e.g., highway=residential). Edits are grouped into immutable changesets, stored in the OSM Planet History, tracking the map’s evolution \cite{item9}. 

This section reviews OSM’s trust management framework (Section 3.1), academic research on data quality assessment and enhancement (Section 3.2), and key insights from trust management systems (Section 3.3). These efforts aim to ensure OSM remains a reliable, real-time resource by identifying trustworthy contributions and refining edits.

\subsection{Trust Management Framework in OSM}
OSM's trust management framework employs a layered approach combining automated rule-based tools, community feedback, and expert intervention to maintain data quality. Automated tools such as the JOSM editor \cite{item26}, OSMOSE \cite{item27}, and Keep Right \cite{item28} proactively flag potential errors using pre-defined rules (e.g., positional inaccuracies, logical inconsistencies, semantic discrepancies), bringing them to the attention of community members who ultimately decide whether to validate or correct these edits. OSMCha \cite{item29} further supports this process by aggregating flagged changesets, offering contributor metadata and highlighting problematic edits for community review. For persistent or malicious issues, the Data Working Group (DWG) \cite{item30} directly resolves disputes or blocks problematic contributors. This collaborative structure ensures swift identification and correction of errors, enhancing trustworthiness and deterring vandalism.

\subsection{Assessing OSM Data Trustworthiness}
Assessing the trustworthiness of OSM data serves two primary purposes within OSM’s trust management framework: it informs users about data trustworthiness, thereby reducing incentives and opportunities for malicious contributions, and it guides community members in prioritising edits for efficient review and correction. Existing methods for assessing OSM trustworthiness typically focus on five key ISO-defined dimensions \cite{item11}: positional accuracy (how precisely mapped features align with real-world locations), thematic accuracy (correctness of attributes such as road classifications or building types), completeness (presence of necessary features and absence of extraneous ones), logical consistency (correct connectivity and adherence to structural rules), and temporal accuracy (map-to-datedness, ensuring data is up-to-date and valid for its intended timeframe). Additionally, trustworthiness assessments consider data reliability, inferred through consensus and consistency across multiple edits over time, and contributor credibility, derived from users’ historical editing patterns and demonstrated expertise. Recent studies have also emphasised the importance of vandalism detection—identifying abnormal edits or malicious activities—to maintain high trust levels in OSM data. 

The following subsections review existing approaches to OSM trustworthiness assessment across these dimensions, noting that studies explicitly addressing thematic accuracy remain scarce due to inherent evaluation complexities.

\subsubsection{Assessing OSM Positional Accuracy}
Studies assessing OSM positional accuracy predominantly employ supervised learning or map-matching techniques. Supervised methods utilise authoritative data to create labelled datasets, training machine learning models to predict positional accuracy based on intrinsic OSM geometric and thematic attributes. For instance, \cite{item31} applies geometric attributes and Random Forest/Lasso regression for evaluating building polygon accuracy, \cite{item32} employs affine or spline transformations to improve building alignment, and \cite{item33} utilises Geographically Weighted Regression (GWR) to model spatial relationships between editing metrics and positional errors.

Alternatively, map-matching directly compares OSM data against authoritative datasets for accuracy estimation without creating training sets, as demonstrated by \cite{item34}, which uses affine transformations and calculates horizontal shifts via RMSE and NSSDA metrics. Both approaches provide valuable frameworks for assessing and enhancing OSM positional accuracy

Limitations: While supervised approaches can predict positional accuracy in real-time, even when authoritative sources are outdated, they heavily rely on the initial availability of authoritative data to create training sets; map-matching methods directly depend on the authoritative dataset's currency, limiting their applicability if reference data is outdated or unavailable.

\subsubsection{Assessing OSM Completeness}
Studies assessing OSM completeness predominantly use map-matching or supervised machine learning methods, along with evidence aggregation techniques. Map-matching approaches directly compare OSM data against authoritative datasets. For instance, \cite{item35} uses geometric matching methods involving authoritative building footprints to quantify completeness by calculating the percentage of matched and unmatched features. Similarly, \cite{item36} assesses completeness by automatically matching OSM land-cover data to satellite-derived land-cover maps.

In comparison to map-matching methods, supervised approaches involve generating labelled training data by aligning OSM data with authoritative references. This labelled data is then utilised to train predictive models. For example, \cite{item37} employs linear regression to model the relationship between OSM building density and completeness from authoritative references, while \cite{item38} employs a supervised machine learning model(Random Forest regression) trained on labelled data derived from visually interpreted high-resolution satellite imagery, utilizing remote sensing indicators (e.g., nighttime lights, spectral indices) and OSM-derived features (e.g., road network data) to predict building completeness.

Another distinct category involves evidence aggregation methods. As demonstrated in \cite{item39}, these methods utilise intrinsic indicators (such as NDVI indices, OSM features like footpath density, and Points of Interest) without the need to create explicit training datasets. In these cases, completeness is estimated through statistical or heuristic evidence aggregation.

Limitations: Supervised and map-matching methods depend heavily on the availability and currency of authoritative data, while evidence aggregation approaches rely significantly on domain expertise to assign appropriate weights to different evidence types.

\subsubsection{Assessing OSM Logical Consistency}
Studies assessing logical consistency in OSM primarily employ rule-based methods using predefined expert guidelines. For instance, \cite{item40} proposes a rule-based methodology that evaluates logical consistency, focusing on internal topology (e.g., non-overlapping buildings, proper road intersections) and external topo-semantic constraints (e.g., cafes must be within buildings, bus stops outside roads). These constraints are validated based on spatial relationships and semantic tags (``Type'' tags). Violations are detected using an ArcGIS tool that calculates error percentages and visually presents results through a web mapping application at regional, city, and feature levels. Results highlight consistency variations among six European cities, noting higher consistency in Utrecht and lower in Athens.

Limitations: Rule-based approaches can overlook errors that don't match predefined constraints and heavily depend on the comprehensiveness and accuracy of expert-defined rules.

\subsubsection{Assessing OSM Data Reliability, Temporal Accuracy and Contributor Credibility}
Studies assessing OSM data reliability, temporal accuracy, and contributor credibility predominantly use intrinsic, rule-based approaches. For instance, \cite{item10} presents an intrinsic method using historical edits to assess all three dimensions. Reliability is evaluated based on edit frequencies and consistency among contributors, while temporal accuracy is calculated through node creation dates and update frequencies. Contributor credibility is inferred from historical editing behaviour and the diversity of edits. Similarly, \cite{item9} computes reliability through direct effects (geometric and thematic consistency), indirect effects (topological relationships), and temporal persistence of edits, alongside contributor credibility from the quality of previous edits. Another study \cite{item41} estimates reliability by analysing spatial and thematic similarity between successive edits, inferring credibility from minimal peer modifications.

In contrast, unsupervised methods, such as \cite{item42}, employ clustering techniques to group contributors based on their editing activity patterns, indirectly assessing credibility without requiring labelled data. However, unsupervised learning algorithms may produce many false positives and false negatives, as they often assume that outliers inherently indicate vandalism—an assumption not always valid within volunteer-based cartography contexts.

Limitations: Rule-based approaches may oversimplify error detection and fail to differentiate clearly between malicious actions and genuine novice mistakes, potentially reducing trust management effectiveness.

\subsubsection{Detecting Vandalism}
Studies assessing vandalism in OSM typically utilise supervised learning methods, often incorporating advanced machine learning algorithms and embedding techniques. For instance, \cite{item43} employs user embeddings derived from a word2vec skip-gram model trained on users' historical editing patterns (shared edits, semantic/temporal sequences), integrating these embeddings as features into a Gradient Boosting Decision Tree (GBDT) classifier. Training labels originate from a manually labelled corpus of vandalised edits from OSMCha and legitimate edits from Daylight Map, enabling binary classification (vandalised or regular). Similarly, \cite{item44} introduces Ovid, a supervised neural network model employing multi-head attention mechanisms to identify potentially vandalised edits within OSM changesets. Inputs include detailed changeset attributes and historical user activity data, while training data comes from community-flagged vandalism in OSM revision histories. Another approach, ``OSMWatchman'' \cite{item45}, uses a supervised Random Forest classifier trained on intrinsic features—such as geometric anomalies and user behaviour patterns—derived from manually introduced artificial vandalism. Target outputs in all these methods are binary, classifying edits as vandalism or legitimate contributions. 

Limitations: Supervised methods are constrained by their dependence on manually labelled data, limiting scalability and comprehensive coverage, and rendering them ineffective in detecting novel or previously unseen patterns of vandalism.

\subsection{Insights from OSM Trust Management}
\subsubsection{Strengths of Established Practices}
OSM benefits significantly from community-driven oversight, embodying the “many eyes” principle, where engaged users swiftly detect and correct errors. Tools such as OSMCha, OSMOSE, Keep Right, and JOSM effectively support community oversight by flagging potential issues for review, allowing experienced mappers to concentrate on complex tasks. The open data model of OSM—with immutable edits stored as publicly accessible changesets—provides transparency, facilitating decentralised data management and historical tracking.

\subsubsection{Challenges and Limitations}
Despite these strengths, OSM’s trust assessment faces several limitations. Contributor credibility, essential for trust assessment, lacks a standardised reputation metric. Although OSMCha presents users' total edits and flags past malicious activities, its predominantly rule-based approach struggles to capture nuanced or evolving vandalism patterns comprehensively. This approach may overlook certain inaccuracies or mistakenly identify valid contributions as suspicious. Moreover, inconsistent global coverage results in disparities between densely edited urban areas and sparsely edited rural or disaster-prone regions, weakening reliability assessments where fewer contributors are available. Additionally, the manual validation required by flagged edits often exceeds volunteer capacity, leading to prolonged unresolved issues. Lastly, concentrated governance among a limited group of influential users may introduce biases, potentially affecting the decentralised ethos and perceived trustworthiness of the platform.	

\subsubsection{Opportunities for Advancement}
OSM could significantly benefit from integrating advanced deep learning and machine learning methods to holistically estimate trustworthiness using all relevant dimensions—positional accuracy, completeness, thematic accuracy, logical consistency, temporal accuracy, data reliability, and contributor credibility—and provide associated uncertainty quantifications. Such comprehensive and transparent trustworthiness scores would help users confidently evaluate the reliability of individual data features, especially in critical scenarios like disaster response or infrastructure planning. Publicly available contributor reputation scores would also encourage mappers to enhance their contributions. By coupling these sophisticated trust assessments and uncertainty measures with transparent and participatory governance, OSM can mitigate risks of elite dominance, promoting a more resilient and trustworthy mapping platform.

\section{Trust Management in Wiki Ecosystems: Practices and Emerging Solutions}

Collaborative wiki platforms \cite{item46}, particularly Wikipedia and Wikidata, have significantly transformed how knowledge is created and disseminated, relying on open participation. While this openness enriches content through diverse contributions, it also poses challenges to accuracy, reliability, and accountability. To maintain trust, Wikipedia and Wikidata employ rigorous editorial policies, community-driven oversight, and automated tools designed to identify vandalism and problematic edits.

This section reviews trust management practices within these platforms. Section 4.1 outlines current mechanisms such as decentralised governance, automated bots, and user reviews. Section 4.2 examines emerging solutions from recent literature, including advanced machine learning techniques, improved reference quality assessments, and governance enhancements. Finally, Section 4.3 summarises key insights, highlighting strengths, limitations, and opportunities for future advancement.

\subsection{Wiki Ecosystems: Current Trust Management Practices}
The Wikimedia ecosystem, administered by the Wikimedia Foundation, encompasses multiple collaborative knowledge-sharing platforms \cite{item47}. Key projects include Wikipedia, a multilingual encyclopedia offering narrative articles, and Wikidata, a structured database providing standardised data for diverse applications. Supporting initiatives include Wikimedia Commons (media repository), Wiktionary (dictionary definitions), and Wikisource (digital texts). These platforms promote openness, transparency, and collective validation, inviting global participation through free licensing.

Trust management within Wikimedia employs a decentralised yet structured approach, combining community-driven oversight, rigorous editorial policies, and automated tools \cite{item47}. Wikipedia enforces verifiability through strict policies, including ``No Original Research'' and ``Neutral Point of View'', requiring contributors to cite reliable sources. Disputes are resolved transparently through community discussions on article talk pages.

Automated and semi-automated tools significantly enhance data quality. Bots like ClueBot NG \cite{item48} rapidly detect vandalism using predefined heuristics and machine learning models trained on historical edits, flagging and reverting suspicious changes and alerting users for human verification. User scripts such as Twinkle \cite{item49} and Huggle \cite{item50} streamline tasks by enabling rapid vandalism reversal and user notifications based on community-defined criteria. ORES \cite{item51}, another critical component, provides probabilistic assessments of edit quality, employing machine learning classifiers (e.g., logistic regression, random forests) trained on historical edit patterns, which help prioritise human review efforts.

Community oversight remains essential, with active contributors maintaining watchlists and monitoring changes \cite{item52}. Users awarded administrator and bureaucrat roles hold advanced privileges—such as page protection and user permission management—assigned after rigorous community vetting. Human judgment ultimately finalises decisions on contentious edits, balancing automated interventions with contextual evaluation.

Although effective, these mechanisms face challenges from the sheer volume of edits, potential false positives, and subtle misinformation risks \cite{item47}. Overall, the Wikimedia ecosystem successfully combines editorial guidelines, automated tools, and community oversight, maintaining robust trust management despite ongoing challenges inherent in volunteer-driven content curation.

\subsection{Emerging Trust Management Solutions in Wiki Ecosystems}
Recent literature proposes several innovative policies and algorithms to enhance content trustworthiness within wiki ecosystems, particularly Wikipedia and Wikidata. Proposed strategies include employing digital trust indicators, such as badges highlighting editor expertise and article quality, alongside collaborative editing policies and clear roles for trusted community members like administrators \cite{item53}. Another approach introduces visual ``Trust-o-meters'' near article titles, displaying metrics such as reference credibility, editor consensus, completeness, and editing stability to transparently convey content reliability \cite{item54}.

AI-driven citation improvement systems have been developed, utilising neural networks (e.g., RoBERTa) and semantic search methods (BM25, BERT) to identify weak citations and recommend better alternatives, with human editors validating these suggestions \cite{item55}. Automated Information Quality assessment methods based on Google's E-A-T model (Expertise, Authority, Trustworthiness) employ web-scraped attributes and machine learning classifiers (Decision Trees, Logistic Regression, K-Means) to efficiently evaluate article quality, emphasising authoritative external citations \cite{item56}.

A hybrid reference quality assessment approach combines Reference Need (RN), measured via a FastText classifier identifying uncited content, and Reference Risk (RR), quantifying misinformation risks from unreliable sources, advocating collaborative editing by pairing novice and experienced contributors \cite{item57}. Wikidata-focused methods integrate crowdsourced evaluations with machine learning algorithms (Random Forest, XGBoost, Gaussian Naive Bayes, SVM, Neural Networks) to assess reference relevance, authoritativeness, and accessibility across multiple languages \cite{item58}. Another two-step approach incorporates initial human assessments followed by automated machine learning predictions (Random Forest, SVM, Naive Bayes) to systematically identify problematic references \cite{item59}.

Structured validation frameworks for Wikidata use community-driven deletion records, deprecation flags, and semantic constraint checks via the Knowledge Graph Toolkit, alongside preventive real-time editor alerts, standardised formatting, and community interfaces for corrections \cite{item60}. Additionally, organisational analyses suggest ``self-organising bureaucratisation'' through community-developed rules and roles as effective in enhancing accountability and mitigating concentrated power structures, thereby increasing content trustworthiness \cite{item61}.

Collectively, these solutions—including visual trust indicators, AI-enhanced citation systems, automated quality assessments, structured validation methods, and robust community governance—offer comprehensive, scalable strategies for ensuring trustworthy content within wiki ecosystems.

\subsection{Insights from Wiki Ecosystem Trust Management}
\subsubsection{Strengths of Established Practices}
\begin{itemize} 
    \item \emph{Community-Driven Governance.} 
    Wikipedia employs a decentralised, consensus-based governance model. Core policies such as ``No Original Research'' and ``Neutral Point of View'' ensure content remains objective and verifiable. Transparent discussions on article talk pages facilitate public resolution of disputes, discouraging biased contributions and promoting reliable content. 
    
    \item \emph{Automated and Semi-Automated Tools.} 
    Bots like ClueBot NG effectively detect and revert vandalism, while semi-automated utilities such as Twinkle and Huggle help volunteers efficiently review changes. Machine learning tools like ORES prioritise edits for human review, combining algorithmic detection with human oversight to maintain content integrity. 
    
    \item \emph{Role-Based Privileges and Continuous Monitoring.} 
    Experienced contributors earn roles (e.g., administrators, bureaucrats) with additional privileges, such as blocking disruptive users or protecting sensitive articles. Watchlists further enable continuous monitoring, encouraging collective responsibility for content accuracy.
\end{itemize}

\subsubsection{Limitations and Ongoing Challenges}
\begin{itemize}
    \item \emph{Power Concentration and Elite Dominance.} 
    Despite decentralisation efforts, power often concentrates among veteran editors and administrators, creating a risk of bias and reduced objectivity in content and policy decisions. 
    
    \item \emph{Human Review Bottlenecks.} 
    While automation assists moderation, nuanced edits still require human judgment. High edit volumes can overwhelm volunteers, increasing burnout risks and allowing subtle misinformation or errors to persist. 
    
    \item \emph{Inconsistent Coverage and Expertise.} 
    High-traffic articles receive consistent scrutiny, while niche or less popular content may remain inadequately reviewed, leaving inaccuracies or biases unchecked due to limited expert engagement. 
    
    \item \emph{False Positives and Missed Edits.} 
    Automated tools occasionally produce false positives or miss sophisticated misinformation, as they lack full contextual understanding. Machine learning models require frequent updates to maintain effectiveness against evolving misinformation tactics. 
    
    \item \emph{Volunteer Burnout and Conflict Resolution Challenges.} 
    Heavy reliance on unpaid volunteers can lead to emotional fatigue from prolonged disputes or ideological conflicts. Existing dispute resolution mechanisms are often slow, risking reduced volunteer participation and oversight.
\end{itemize}

\subsubsection{Opportunities for Future Advancement}
\begin{itemize}
    \item \emph{Enhanced Machine Learning Tools.} 
    Incorporating advanced natural language processing and deep learning into existing moderation tools (e.g., ORES) could improve context sensitivity and accuracy. Continuous human-in-the-loop feedback systems can further reduce false positives and adapt to emerging misinformation tactics. 
    
    \item \emph{Transparent Reputation Metrics.} 
    Developing standardised and transparent metrics to assess contributor reliability—such as visual ``Trust-o-meters'' indicating article quality or badges highlighting editor expertise—could incentivise high-quality contributions and enhance user trust. 
    
    \item \emph{AI-Assisted Trust Indicators.} 
    Introducing systems that clearly display trustworthiness scores with accompanying explanations and uncertainty indicators could foster transparency, accountability, and improved content sourcing practices. 
    
    \item \emph{Specialised Expert Groups.} 
    Establishing dedicated task forces or niche patrols, possibly in collaboration with academic institutions or professional societies, could ensure under-monitored or specialised topics receive adequate expert attention, addressing coverage gaps. 
    
    \item \emph{Improved Conflict Resolution Processes.} 
    Streamlining dispute resolution through semi-automated tools that summarise debates, highlight consensus, and identify contentious points could expedite resolutions, relieve volunteer burdens, and prevent content stagnation. 
    
    \item \emph{Decentralised and Distributed Governance.} 
    Implementing decentralised governance methods—such as rotating leadership, community-elected panels, or blockchain-based voting—could help mitigate power concentration, distribute influence more evenly, and enhance overall content.
\end{itemize}

\section{Trust Management in Social Media: Practices and Emerging Solutions}

Social media platforms, engaging billions of global users, significantly shape public discourse and provide valuable crowd-generated data for real-time insights, such as breaking news and social movements \cite{item62}. However, their openness also creates vulnerabilities to misinformation, spam, and malicious activities, making trust management a critical challenge \cite{item63}.

X and YouTube stand out due to their large user bases and extensive content reach \cite{item64}. X facilitates rapid, text-based discussions, whereas YouTube offers video-driven narratives capable of presenting complex information. 

Sections 5.1 and 5.2 discuss the current trust management practices of X and YouTube, respectively. Literature-proposed enhancements to these platforms are reviewed in Section 5.3. Finally, Section 5.4 synthesises insights regarding the strengths, limitations, and future advancement opportunities in social media trust management.

\subsection{Trust management in X: Current Practices and Approaches}
X employs a multi-layered trust management system integrating automated tools and user feedback \cite{item65}. Machine learning models (e.g., logistic regression, random forests, deep learning like convolutional or transformer-based architectures) detect spam, bots, and abusive content by analysing posts, user behaviour, and network patterns, flagging violations for labelling, visibility reduction, or removal \cite{item66}. Rule-based filters complement these by swiftly blocking content with banned keywords or spam-like phrases, updated rapidly by policy teams to counter misinformation. Users report harmful posts, triggering human moderator reviews, balancing automation with nuanced oversight.

X’s Community Notes (formerly Birdwatch) enables eligible users—screened by account longevity, rule compliance, and phone verification—to add referenced clarifications to misleading posts \cite{item67}. Notes, initially visible to annotators, go public after diverse reviewers rate them “helpful” via a “bridging” method to reduce bias, though eligibility criteria may limit diverse input. Historically, X’s blue verification badge signalled trust, but the 2022 X Premium subscription shifted this to a paid model with perks like editing and fewer ads, not ensuring identity verification or Community Notes access \cite{item68}. Critics note this weakens trust, risking impersonation despite legacy badge distinctions.

X refines rules, enforcement, and tools continuously, combining machine learning, filtering, and Community Notes \cite{item69}. Yet, challenges remain: Community Notes’ restrictions may exclude perspectives, subscription verification may elevate unreliable voices, and machine learning/filtering struggles with regional nuances and evolving misinformation due to language and behaviour complexity \cite{item70}, \cite{item71}. These gaps highlight the need for adaptive trust strategies to balance discourse and harm mitigation.

\subsection{Trust Management in YouTube: Current Practices and Approaches}
YouTube’s multi-layered trust management framework integrates community guidelines, algorithmic analysis, and human moderation to curb harmful or misleading content \cite{item72}. Its guidelines ban hate speech, harassment, misleading health claims, and extremist propaganda, with flagged videos facing removal, channel strikes, or termination. Machine learning models analyse metadata (titles, descriptions, tags), audio via speech recognition, and visuals through image/object detection to spot violations like spam or violence \cite{item73}, \cite{item74}. The Content ID system matches uploads against a registered database to enforce copyright, targeting piracy over misinformation.

Demonetization withdraws ads from videos unfit for advertisers, incentivising creators to maintain accuracy and quality, even for borderline content not warranting removal \cite{item75}. User engagement (likes, comments, watch time) shapes recommendation algorithms, impacting visibility. Human moderators and contractors review flagged content, applying guidelines and local laws, with user appeals triggering further reviews to address context, nuance, and evolving misinformation \cite{item76}. Critics highlight challenges with moderating 500+ hours of uploads per minute and inconsistent demonetization, yet YouTube balances automated detection, financial penalties, and human oversight to reduce untrustworthy content while supporting creators.

\subsection{Emerging Trust Management Solutions in Social Media}
Recent literature proposes policies and algorithms to enhance trust management on X and YouTube, categorised as: (1) policy-driven recommendations, (2) rule-based and evidence aggregator algorithms, (3) supervised and hybrid machine learning algorithms, and (4) unsupervised and graph-based machine learning algorithms.
\begin{itemize} 
    \item \emph{Policy-driven Recommendations.} 
    Policy solutions focus on human and platform factors affecting trust. On YouTube, creator reputation, content quality, and social validation are key \cite{item77}. Recommendations include transparent labelling, thumbnails, descriptions, plus digital literacy, critical thinking, and stricter verification guidelines. 
    
    \item \emph{Rule-based and Evidence Aggregator Algorithms.} 
    Heuristic systems include an evidence aggregator for X posts, assessing readability, completeness, usefulness, and trustworthiness using Natural Language Processing (NLP) sentiment tools \cite{item78}, and T-CREo, a real-time rule-based framework evaluating X post credibility across textual, user, and social dimensions \cite{item79}. 
    
    \item \emph{Supervised and Hybrid Machine Learning Algorithms.} 
    Supervised and hybrid methods classify misinformation using text and metadata. A supervised system integrates user reputation, activity, and weighted feature-ranking with classifiers (Random Forests, Decision Trees, Naïve Bayes, FR\_NB) \cite{item80}. A hybrid approach blends supervised (Random Forest, Logistic Regression) and semi-supervised 'agreement-based retraining' for iterative accuracy gains \cite{item81}. For crises, a supervised pipeline uses bigram TF-IDF and metadata (followers, retweets, verification) with SVM, Random Forest, XGBoost, AdaBoost, Decision Trees, MLPs, and k-NN \cite{item82}. YouTube misinformation detection employs NLP-based supervised classifiers (SVC, Random Forest, ExtraTrees, XGBoost, AdaBoost) on captions \cite{item83} and Transformer models (BERT, RoBERTa, ELECTRA, MPNet) with sliding-window for transcripts \cite{item84}. A statistical-supervised hybrid scores credibility via propagation and profiles, using Naïve Bayes, Decision Trees, Random Forest, and Gradient Boosting (CatBoost, LightGBM, XGBoost) \cite{item85}. 
    
    \item \emph{Unsupervised and Graph-based Machine Learning Algorithms.} 
    Unsupervised and graph-based methods detect anomalies and assess credibility. An iterative trust propagation algorithm compares X posts to credible sources using semantic, spatial, and temporal measures \cite{item86}. An OCC framework with SSVDD and graph regularisation classifies untrusted users via metadata and graph structure \cite{item87}. A hybrid 'Believability' metric uses unsupervised LINE embedding and supervised GRU-based classification to spot rumour spreaders \cite{item88}. 
    
    \item \emph{Benchmark Datasets and Evaluation Tools.} 
    TwiBot-20, a benchmark dataset for X bot detection, supports testing supervised (Random Forests, LSTM, GCN) and unsupervised/rule-based methods, aiding trust enhancements \cite{item89}.
\end{itemize}

\subsection{Insights from Social Media Trust Management}
\subsubsection{Strengths of Established Practices}
\begin{itemize} 
    \item \emph{Hybrid Moderation Models.} 
    Platforms like X and YouTube effectively integrate AI-driven tools (machine learning classifiers, rule-based filters) with human oversight (user reporting, moderators). This hybrid approach enables efficient handling of massive content volumes, ensuring rapid detection while retaining nuanced human judgment for complex scenarios. 
    
    \item \emph{Clear Policy Frameworks.} 
    Explicit community guidelines combined with transparent enforcement policies—warnings, suspensions, bans—establish consistent behavioural standards, facilitating reliable moderation decisions and content integrity. 
    
    \item \emph{Real-Time Community Engagement.} 
    User feedback systems, such as X’s reporting features and YouTube’s comment moderation tools, actively involve users in platform governance, quickly identifying emerging misinformation trends and complementing automated moderation. 
    
    \item \emph{Trust Incentivisation Mechanisms.} 
    Platforms implement initiatives like X’s Community Notes (crowdsourced contextual annotations) and YouTube’s monetisation programs to incentivise credible contributions. Such systems align user interests with maintaining high content standards.
\end{itemize}

\subsubsection{Challenges and limitations}
\begin{itemize} 
    \item \emph{Overwhelming Content Volume.} 
    The immense volume and velocity of user-generated content (millions of posts and hours of videos uploaded daily) often outpace moderation capabilities, allowing misinformation to spread rapidly and evade timely intervention. 
    
    \item \emph{Balancing Expression and Safety.} 
    Platforms face persistent challenges in achieving a balance between freedom of speech and effective harm reduction, intensified by regional, cultural, and political differences, as well as evolving misinformation tactics exploiting policy gaps. 
    
    \item \emph{Algorithmic Bias and Lack of Transparency.} 
    Automated moderation systems may inadvertently reinforce biases or misclassify legitimate content, disproportionately affecting specific user groups. Additionally, proprietary algorithms remain largely opaque, leading to concerns regarding fairness, accountability, and susceptibility to manipulation. 
    
    \item \emph{Monetisation and Credibility Conflicts.} 
    Policies like demonetization or subscription-based verification can disproportionately disadvantage smaller creators or marginalised voices while inadvertently allowing bad actors to purchase credibility. This misalignment risks undermining user trust in platform integrity. 
    
    \item \emph{Concentration of Authority.} 
    Despite decentralisation efforts, moderation authority tends to concentrate among influential creators or internal moderation teams, introducing systemic biases and vulnerability points that could compromise the overall trustworthiness of platform content.
\end{itemize}
\subsubsection{Opportunities for Future Advancement}
\begin{itemize} 
    \item \emph{Transparent and Explainable AI Scoring.} 
    Deploying AI systems with built-in uncertainty estimates and publicly accessible explanations can improve moderation transparency and reduce false positives. Platforms can adopt adaptive moderation that dynamically learns from community feedback, adjusting scoring algorithms when community-provided evidence contradicts initial AI assessments. 
    
    \item \emph{Decentralised Governance Structures.} 
    Implementing rotating moderation panels and transparent, distributed reputation systems can democratise moderation authority, minimising bias, reducing vulnerability to manipulation, and enhancing collective oversight. 
    
    \item \emph{Advanced Context-Aware Analysis.} 
    Leveraging state-of-the-art NLP and multimodal technologies (such as Transformer-based models) will enable better contextual understanding, significantly reducing misclassification and inappropriate content removals. 
    
    \item \emph{Proactive User Empowerment Tools.} 
    Developing mechanisms to proactively notify users before content removal allows them to clarify context or revise content, fostering fairness and user engagement. 
    
    \item \emph{Cross-Platform Evaluation Standards.} 
    Establishing standardised, transparent benchmarks for trust management performance can ensure consistent moderation outcomes across platforms and diverse cultural contexts, facilitating continuous improvement and equitable content governance.
\end{itemize}

\section{Trust Management in Mobile Crowdsensing Platforms}

Mobile crowdsensing platforms utilise smartphones and portable sensors to collect real-time environmental data, expanding upon traditional monitoring methods \cite{item90}. Effective trust management—including validating submissions, identifying anomalies, and incentivising high-quality contributions—is essential to maintain dataset reliability. This section reviews current trust management practices in five prominent platforms: Waze (Section 6.1), OpenSignal (Section 6.2), Sensorly (Section 6.3), Safecast (Section 6.4), and AirCasting (Section 6.5). Section 6.6 synthesises key insights regarding strengths, limitations, and opportunities for advancing trust management strategies within these platforms.

\subsection{Trust Management in Waze}
Waze \cite{item43}, \cite{item91} ensures accurate, real-time traffic data through reputation systems, community governance, and algorithmic validation:
\begin{itemize}
    \item \emph{Reputation and Community Structure.} 
    Contributors earn points and ranks (``Baby Wazer'' to ``Royalty Wazer'') by reporting hazards, confirming updates, or editing maps \cite{item92}. Experienced users advance to roles such as Area Managers or Regional Coordinators, with enhanced map-editing privileges and responsibilities for resolving disputes and training new editors. 
    
    \item \emph{Technical Validation.} 
    Waze employs rule-based and machine learning approaches, cross-verifying user reports against anonymised location and speed data to flag anomalies \cite{item93}. Automated routines validate map edits for geometry accuracy, duplicates, and conflicts. 
    
    \item \emph{Selective Transparency.} 
    Users see their own ranks, and active editors see collaborators’ levels during map editing. Casual users only view milestone badges, preventing competitive misuse or spam.
\end{itemize}

This integrated framework mitigates misinformation, rewarding reliability and sustaining user confidence in navigation data.

\subsection{Trust Management in OpenSignal}
OpenSignal collects anonymised, crowdsourced cellular network data (signal strength, speed, coverage) from millions of devices, focusing on validation and aggregation \cite{item94}, \cite{item95}:
\begin{itemize} 
    \item \emph{Passive Data Collection.} 
    The app automatically logs cellular signals, GPS locations, and device details post-consent, minimising human error and providing consistent real-world measurements. 
    
    \item \emph{Multi-layered Validation.} 
    Combines rule-based filters (e.g., removing impossible readings) with machine learning-based anomaly detection to identify suspicious data points and assign them lower confidence. 
    
    \item \emph{Robustness Through Redundancy.} 
    Aggregating overlapping measurements from multiple devices under similar conditions dilutes errors or malicious contributions. Data from widely used devices and frequently measured locations implicitly gain higher trust over time.
\end{itemize}

These methods ensure accurate network maps useful for both consumers and industry stakeholders.

\subsection{Trust Management in Sensorly}
Sensorly crowdsources cellular and Wi-Fi performance data similarly to OpenSignal, relying on automated collection and anomaly filtering \cite{item96}, \cite{item97}:
\begin{itemize}
    \item \emph{Passive Data Collection.} 
    Continuously logs signal strength and GPS data in the background, reducing user-induced errors. 
    
    \item \emph{Data Validation.} 
    Aggregation establishes baseline network performance, while rule-based filters and statistical or machine learning-based models detect anomalous readings. Outliers—such as improbable signal spikes—are flagged or deprioritised. 
    
    \item \emph{Implicit Trust via Redundancy.} 
    Without public reputation scores, Sensorly uses large-scale data overlap, weighing consistent measurements heavily to maintain accuracy and minimise faulty inputs.
\end{itemize}

This approach ensures reliable network coverage maps by naturally filtering out anomalies.

\subsection{Trust Management in Safecast}
Safecast collects environmental radiation measurements through volunteer-driven, open-source practices initiated post-Fukushima disaster \cite{item98}, \cite{item99}:
\begin{itemize} 
    \item \emph{Open Hardware Philosophy.} Utilises standardised, open-source sensors (e.g., bGeigie) with documented calibration standards, ensuring verifiable hardware and software quality. 
    \item \emph{Cross-verification of Data.} 
    Registered volunteers submit data tagged with device IDs, timestamps, and GPS coordinates. Overlapping measurements are compared to identify significant deviations, prompting further community review or operator contact. 
    
    \item \emph{Transparency and Open Access.} 
    All data is openly accessible under clear licensing terms, linking each measurement to specific devices and contributors. This transparency enables independent verification, discouraging intentional data manipulation.
\end{itemize}

Safecast's commitment to transparent hardware standards, redundant data checks, and public accountability sustains data accuracy and community trust.

\subsection{Trust Management in AirCasting}
AirCasting gathers community-driven air quality data using portable sensors, focusing on transparent, calibration-centric practices \cite{item100}:
\begin{itemize} 
    \item \emph{Open-Source Infrastructure.} 
    Supports diverse, documented sensor kits, enabling peer verification of device quality and measurement accuracy. 
    
    \item \emph{Calibration and Comparative Validation.} 
    Contributors regularly calibrate devices against reference standards, detecting and correcting sensor drift. Multiple concurrent local measurements facilitate the rapid identification and correction of anomalous readings. 
    
    \item \emph{Community Oversight and Transparency.} 
    All collected data appear on open-access maps, inviting scrutiny from researchers and local communities. Open discussions in community forums further address anomalies, discouraging misinformation and reinforcing trust.
\end{itemize}

These combined efforts—open-source design, rigorous calibration, redundancy checks, and transparent sharing—help AirCasting reliably reflect real-world air quality conditions.

\subsection{Emerging Trust Management Solutions in Mobile Crowdsensing Platforms}
Recent literature presents diverse policies and algorithms to enhance data trustworthiness in mobile crowdsensing platforms, addressing reliability, misinformation detection, privacy protection, and incentive fairness.

Blockchain-based solutions have become prominent, such as BlockSense, which introduces a Proof-of-Data (PoD) consensus algorithm using zero-knowledge proofs (zk-SNARK) and homomorphic data perturbation for secure, privacy-preserving verification of sensory data \cite{item14}. This rule-based blockchain approach, supported by smart contract-driven economic incentives, effectively promotes trustworthy data submissions.

Several approaches leverage machine learning and statistical methods. A hybrid solution integrates supervised two-layer Long Short-Term Memory (LSTM) neural networks for device fingerprinting with rule-based similarity evaluations for detecting duplicated or malicious reports \cite{item101}. Another advanced framework, Explainable Fake Sequential Data (XFSD), combines supervised LSTM neural networks with explainable AI methods (SHAP, CAM, LIME) for transparent adversarial data detection \cite{item102}. Additionally, False Sequential Data Detection (FSD) employs supervised LSTM models alongside unsupervised One-Class SVM classifiers for anomaly detection to isolate malicious sensor inputs \cite{item103}.

Privacy-preserving and cryptographic methods also appear extensively. The Privacy-preserving Trust-Aware Group Formation (PTAGF) framework uses heuristic algorithms for trust evaluation and group formation, combined with two-layer encryption and k-anonymity to safeguard participant privacy \cite{item104}. Another cryptography-driven framework employs blind and group signatures with anonymous reputation models and pseudonym management to counter anonymity abuses and ensure data integrity \cite{item105}.

Hybrid statistical and decentralised methods are also notable. A collaborative reputation scoring algorithm combines statistical unsupervised outlier detection with decentralised vote-based heuristics, robustly quantifying user reliability \cite{item106}. Furthermore, an anchor-assisted vote-based method uses highly trusted ``anchor'' users guiding distributed community voting, enhancing misinformation filtering without explicit machine learning techniques \cite{item107}.

Statistical interpolation and optimisation techniques are utilised as well. The Trustworthy and Cost-Effective Cell Selection (TCECS) framework applies unsupervised Ordinary Kriging interpolation to identify trustworthy data, coupled with Regularised Mutual Coherence optimisation to achieve cost-effective sensor placement \cite{item108}.

Human-centric approaches are presented in a statistical trust evaluation framework incorporating subjective user opinions and objective sensor data, employing the Kolmogorov–Smirnov test to handle individual trust dispositions without traditional supervised learning \cite{item90}. Moreover, the Data Trustworthiness Enhanced Crowdsourcing Strategy (DTCS) integrates attribute-based assessments, group division via metagraph theory, unsupervised anomaly detection, and game-theoretic incentives to mitigate internal threats and collusion effectively \cite{item109}.

Collectively, these methodologies-spanning blockchain, hybrid machine learning, statistical analyses, cryptographic, heuristic, and human-centric approaches --- significantly advance trust management capabilities in mobile crowdsensing environments.

\subsection{Insights from Mobile Crowdsensing Trust Management}
\subsubsection{Strengths of Established Practices}
\begin{itemize} 
    \item \emph{Robust Validation Mechanisms.} 
    Platforms like Waze, OpenSignal, and Sensorly effectively combine rule-based algorithms, machine learning, and community oversight for scalable anomaly detection, significantly enhancing data reliability. 
    
    \item \emph{Redundancy and Aggregation.} 
    Mass user participation ensures overlapping data submissions, naturally diluting the impact of individual inaccuracies or malicious reports, thus creating self-correcting datasets. 
    
    \item \emph{Standardised and Passive Data Collection.} 
    Automated background data logging (OpenSignal, Sensorly) minimises human error and standardises measurements, complemented by open-source hardware and clear calibration standards (Safecast, AirCasting). 
    
    \item \emph{Transparent Operational Practices.} 
    Platforms like Safecast and AirCasting publicly share sensor data, hardware schematics, and calibration protocols, fostering community verification, accountability, and data trustworthiness. 
    
    \item \emph{Motivational Elements.} 
    Reputation-based systems, notably in Waze, encourage constructive participation through gamification and structured user advancement, indirectly improving data quality.
\end{itemize}

\subsubsection{Challenges and Limitations}
\begin{itemize}	
    \item \emph{Geographic and User Participation Disparities.} 
    Sparse participation in remote or underserved areas leads to inconsistent data quality and reliability. 
    
    \item \emph{Device Heterogeneity.} 
    Variations in device hardware quality, calibration accuracy, and maintenance levels can cause measurement inconsistencies and complicate anomaly detection. 
    
    \item \emph{Limited Public Reputation Metrics.} 
    Most platforms lack clear public indicators of contributor reliability, limiting the ability of end-users and stakeholders to assess data trustworthiness directly. 
    
    \item \emph{Context Sensitivity Challenges.} 
    Current anomaly detection methods may fail to accurately interpret region-specific conditions, misclassifying legitimate local anomalies or dismissing valid global outliers. 
    
    \item \emph{Vulnerability to Sophisticated Attacks.} 
    Platforms remain susceptible to coordinated misinformation or collusion attacks, particularly where redundancy is insufficient.
\end{itemize}

\subsubsection{Opportunities for Future Advancement}
\begin{itemize}	
    \item \emph{Blockchain-Based Validation.} 
    Adopting blockchain solutions like Proof-of-Data (PoD) with zero-knowledge proofs (zk-SNARK) can enhance secure, transparent, and privacy-preserving data verification \cite{item14}. 
    
    \item \emph{Advanced Machine Learning Integration.} 
    Implementing hybrid supervised LSTM neural networks with explainable AI techniques (e.g., SHAP, CAM, LIME) would enhance transparent, context-aware anomaly detection \cite{item102}. 
    
    \item \emph{Enhanced Privacy Frameworks.} 
    Incorporating cryptographic techniques such as blind signatures, group signatures, and k-anonymity could effectively balance anonymity with accountability \cite{item104}, \cite{item105}. 
    
    \item \emph{Context-Aware Statistical Models.} 
    Employing sophisticated statistical methods like Ordinary Kriging interpolation combined with optimisation techniques (Regularised Mutual Coherence) can accurately detect anomalies considering localised conditions \cite{item108}. 
    
    \item \emph{Hybrid Trust Evaluation Approaches.} 
    Combining decentralised community voting anchored by highly trusted users (anchors) with statistical outlier detection can robustly filter misinformation without intensive computational resources \cite{item107}. 
    
    \item \emph{Adaptive Reputation and Incentive Systems.} 
    Utilising dynamic, adaptive reputation scoring and integrating game-theoretic incentive mechanisms (metagraph theory-based frameworks) would encourage honest data contributions and deter malicious behaviours effectively \cite{item109}.
\end{itemize}
By leveraging these advanced methodologies, mobile crowd-sensing platforms can significantly enhance trust management, addressing current limitations and ensuring sustained high-quality, reliable data contributions.

\section{Trust Management in Specialised Environmental Crowdsourcing Platforms} 

Specialised environmental crowdsourcing platforms involve active citizen participation in documenting and classifying environmental phenomena such as flooding, biodiversity, weather events, and astronomical objects. Unlike crowdsensing platforms (e.g., Safecast or AirCasting) that passively collect sensor-generated data from mobile devices, environmental crowdsourcing relies on deliberate user submissions—typically involving structured observations, visual documentation, or classifications guided by explicit protocols to ensure accuracy and reliability. Sections 7.1 and 7.2 discuss current trust management practices of two prominent environmental crowdsourcing platforms: MyCoast and Galaxy Zoo (Zooniverse). Section 7.3 reviews emerging solutions from the literature designed to enhance trust management. Finally, Section 7.4 synthesises insights from these practices and solutions, identifying strengths, limitations, and future opportunities in trust management for environmental crowdsourcing platforms.
\subsection{MyCoast}
MyCoast is a platform deployed along various US coastlines that encourages citizens to document flooding, storm damage, tidal events, and other coastal impacts by uploading geo-referenced photos and descriptions \cite{item110}, \cite{item111}, \cite{item112}. At its core, MyCoast’s trust management framework blends human validation, expert-defined guidelines, and automated checks to ensure the reliability of crowd-submitted observations. Volunteers must include location data and photographic evidence, allowing local officials and trained project coordinators to compare reported conditions with known problem areas, weather data, and historical event patterns. By requiring a clear visual record whenever possible, MyCoast makes it harder for contributors to post false information and simpler for moderators to spot discrepancies or unrealistic submissions. 

While MyCoast relies significantly on human oversight—municipal staff and researchers review flagged entries—the platform also employs rule-based logic and evolving machine learning techniques \cite{item113}. For instance, user-submitted photos tagged with specific tidal levels or storm categories are cross-referenced with public tide gauges or weather data; anomalies such as reports of ``major flooding'' conflicting with official metrics prompt further scrutiny.

Over time, MyCoast has begun experimenting with image analysis techniques to identify objects or water levels in user-submitted photos, enhancing the platform’s ability to detect mislabeled events or inconsistent measurements \cite{item113}. Expert-defined standards regarding what constitutes minor, moderate, or major flooding likewise guide the classification of reports, reducing ambiguity around user-submitted terms. Beyond these technical safeguards, MyCoast actively fosters a community culture that discourages inaccurate reports. Participants learn how to document events more consistently through in-app prompts or training materials, and the platform highlights successful case studies where accurate crowd data informed local decision-making—thereby reinforcing the value of precise submissions. Contributors who consistently provide well-documented, context-rich observations gain implicit credibility, as local authorities and researchers often rely on their input in subsequent events. By blending clear user guidelines, automated validation tools, and human review, MyCoast produces data that local governments, scientists, and the broader public can trust to guide responsive coastal management.
\subsection{Galaxy Zoo (Zooniverse)}
Galaxy Zoo, a key Zooniverse project launched in 2007, engages volunteers to classify galaxy shapes, with each image labelled by $>40$ users for consensus \cite{item114}, \cite{item115}. New contributors are trained via tutorials and guides, ensuring standardised classification \cite{item58}. Redundancy and aggregation (vote-count or decision-tree consensus) dilute outliers, yielding reliable decisions with uncertainty estimates; images require $10-40+$ classifications before retirement to avoid spurious results \cite{item116}.

Accuracy and credibility are enhanced by community oversight and algorithms. Consensus algorithms with user-weighting reduce the influence of inconsistent classifiers (~5\% with near-zero weight), filtering guesses without bans, while prioritising consistent contributors \cite{item117}. ZooBot AI pre-screens images, auto-labelling simple cases and flagging complex ones for humans, improving accuracy and AI training in a human-in-the-loop model \cite{item114}. Community features like Talk boards foster reliable participation—volunteers discuss, get expert feedback, and spot errors, with experienced users mentoring and flagging spam alongside automated detection \cite{item115}. Recognition (e.g., publication credits) motivates quality contributions. These elements—consensus validation, algorithmic checks, and volunteer accountability—ensure accurate, reliable, and credible classifications \cite{item118}.

\subsection{Emerging Trust Management Solutions in Specialised Environmental Crowdsourcing} 
Recent literature proposes algorithms and policies to enhance trust in environmental crowdsourcing platforms by detecting untrustworthy submissions and identifying reliable contributions, including semi-supervised learning \cite{item119}, adaptive anomaly detection \cite{item120}, credibility models, and rule-based uncertainty handling \cite{item121}.

A semi-supervised approach \cite{item119} uses a deep CNN (ResNet50) with Adaptive Clustering (AC) and Entropy Separation (ES) to align features and distinguish known categories from anomalies. It leverages labelled and unlabeled images to classify environmental observations accurately and isolate unreliable instances, boosting platform reliability.

Another method \cite{item121} manages uncertainty in citizen classifications via a rule-based statistical process: normalising scores, reallocating 'Don't Know' responses, and boosting confidence with consensus. It refines vote distributions and uncertainty measures, enhancing accuracy and credibility of environmental data.

A VGI-focused credibility model \cite{item120} for flood crises integrates Getis-Ord Gi* spatial clustering and fuzzy logic to assess geographic consistency and generate hazard credibility scores from incident reports, coordinates, elevation, and volunteer locations. This rule-based and unsupervised approach distinguishes reliable from less credible data, aiding disaster monitoring.

Together, these scalable solutions strengthen trust management, improving the quality and reliability of environmental crowdsourcing by countering misinformation effectively.

\subsection{Insights from Specialised Environmental Crowdsourcing Trust Management}
\subsubsection{Strengths of Established Practices}
\begin{itemize} 
    \item \emph{Hybrid Validation Mechanisms.} 
    Platforms such as MyCoast and Galaxy Zoo effectively integrate human validation, automated rule-based algorithms, and machine learning techniques, ensuring robust detection of anomalies and misinformation. 
    
    \item \emph{Structured Contribution Protocols.} 
    Clear guidelines, structured submission processes, and explicit training (as in Galaxy Zoo) enhance contributor reliability and improve overall data accuracy and consistency. 
    
    \item \emph{Community Oversight and Engagement.} 
    Active community forums, expert mentorship, and consensus-driven validations (Galaxy Zoo) foster collective accountability, reinforcing contributor motivation and enhancing data credibility. 
    
    \item \emph{Visual and Georeferenced Evidence.} 
    Platforms like MyCoast emphasise photo documentation and geo-referencing, making submissions easily verifiable and harder to falsify, thus significantly boosting trustworthiness.
\end{itemize}

\subsubsection{Challenges and Limitations}
\begin{itemize} 
    \item \emph{Variability in Contributor Expertise.} 
    Varying levels of participant skill and experience can introduce inconsistent data quality, requiring continuous training and oversight. 
    
    \item \emph{Algorithm Transparency Issues.} 
    Proprietary or undisclosed validation algorithms might limit user trust or understanding of data filtering processes, potentially reducing engagement. 
    
    \item \emph{Limited Contributor Reputation Metrics.} 
    The absence of explicit reputation systems can make it challenging for platforms to differentiate between consistently reliable contributors and occasional or inaccurate participants. 
    
    \item \emph{Contextual and Regional Differences.} 
    Algorithms and validation mechanisms may not adequately accommodate local environmental nuances or context-specific anomalies, risking misclassification or overlooking genuine observations.
\end{itemize}

\subsubsection{Opportunities for Future Advancement}
\begin{itemize} 
    \item \emph{Advanced Semi-Supervised Learning Techniques.} 
    Leveraging adaptive clustering, entropy separation, and universal domain adaptation frameworks can significantly enhance anomaly detection and cross-dataset reliability \cite{item119}. 
    
    \item \emph{Systematic Uncertainty Handling.} 
    Implementing rule-based statistical methods to systematically manage and clarify user-generated uncertainties can markedly improve classification accuracy and credibility \cite{item121}. 
    
    \item \emph{Spatial Credibility Assessment Models.} 
    Integrating geographic clustering and fuzzy logic-based credibility scoring, especially during disaster scenarios like flooding, can robustly differentiate reliable from unreliable submissions \cite{item120}. 
    
    \item \emph{Adaptive Contributor Reputation Systems.} 
    Introducing dynamic contributor scores based on historical accuracy, consistency, and community validations can incentivise sustained high-quality participation and facilitate better trust management. 
    
    \item \emph{Enhanced Community-Algorithm Interaction.} 
    Combining community-driven moderation with AI-driven pre-screening (similar to Galaxy Zoo's ZooBot) can efficiently manage data volumes, focusing human efforts on ambiguous or complex cases, further refining trust assessments.
\end{itemize}

By adopting these advanced methodologies and refining current strengths, environmental crowdsourcing platforms can significantly enhance trust management, ensuring sustained, high-quality, and credible environmental data.

\section{Trust Management in Specialised Review Crowdsourcing Platforms}

Consumer review platforms like Yelp, TripAdvisor, and Amazon heavily influence consumer decisions through user-generated feedback, necessitating robust trust management to combat fraudulent reviews. Sections 8.1–8.3 discuss current trust management practices of these platforms. Section 8.4 reviews literature-proposed enhancements, and Section 8.5 synthesises key insights.

\subsection{Yelp’s Trust Management Practices}
Yelp employs a comprehensive approach combining automated review filtering, community moderation, user transparency, and business engagement \cite{item122}, \cite{item123}.
\begin{itemize} 
    \item \emph{Automated Filtering.} 
    Yelp uses proprietary machine learning algorithms to detect suspicious review patterns based on user history, content quality, and behavioural indicators (e.g., coordinated review activity). Reviews flagged as questionable are marked as ``not currently recommended'', accessible but less visible to users. 
    
    \item \emph{Community Oversight.} 
    Clear community guidelines prohibit spam, conflicts of interest, and promotional content. Users can flag questionable reviews, triggering a moderation process involving both automated checks and human review. 
    
    \item \emph{User Reputation.} 
    Yelp encourages detailed user profiles and recognises consistent, high-quality contributors through its ``Elite Squad'' program, boosting their credibility and incentivising authentic participation. 
    
    \item \emph{Business Responses.} 
    Verified business owners can publicly respond to reviews, adding transparency and allowing users to evaluate the legitimacy of claims.
\end{itemize}

Yelp maintains a strategic balance between transparency and algorithmic secrecy to protect its system from manipulation, occasionally drawing criticism from businesses over potentially suppressed legitimate reviews.

\subsection{TripAdvisor’s Trust Management Practices}
TripAdvisor integrates community-driven reporting, algorithmic detection, and transparent user and business interactions \cite{item124}.
\begin{itemize} 
    \item \emph{Algorithmic and Community Checks.} 
    Proprietary algorithms identify suspicious patterns (e.g., sudden spikes, repetitive phrasing). Users can report violations, leading moderators to reassess and possibly remove flagged reviews. 
    
    \item \emph{Reviewer Credibility.} 
    Reviewers' histories and awarded badges (e.g., ``Senior Reviewer'') enhance transparency, helping readers gauge reviewer reliability. 
    
    \item \emph{Business Engagement.} 
    Property owners can publicly respond to reviews, providing context and accountability, which assists readers in evaluating credibility. 
    
    \item \emph{Ranking Algorithms.} 
    Established reviewers' feedback significantly influences rankings, reducing susceptibility to ``review bombing'' and spam.
\end{itemize}

TripAdvisor continuously updates detection systems, introducing adaptive measures (e.g., verifying health and safety measures) to counter evolving manipulation tactics.

\subsection{Amazon’s Trust Management Practices}
Amazon safeguards review integrity through verification indicators, automated monitoring, structured reviewer programs, and active moderation \cite{item182}.
\begin{itemize} 
    \item \emph{Verified Purchase Indicators.} 
    Reviews from confirmed buyers receive a ``Verified Purchase'' label, helping distinguish authentic consumer feedback from potentially fraudulent submissions. 
    
    \item \emph{Machine Learning-Based Detection.} 
    Proprietary algorithms detect suspicious patterns such as rapid review postings or repeated IP address submissions, potentially leading to review removal or reduced visibility. 
    
    \item \emph{Community Moderation.} 
    Users report inappropriate or deceptive reviews, initiating moderation processes combining automated systems and human oversight. 
    
    \item \emph{Structured Reviewer Programs.} 
    Programs like Amazon Vine invite trusted users to review products, enforcing strict guidelines and transparency requirements to maintain objectivity. 
    
    \item \emph{Highlighting Trustworthy Reviews.} 
    Amazon prominently displays helpful, verified reviews based on user votes and quality assessments to guide consumer decisions effectively.
\end{itemize}
Amazon continuously refines these practices, incorporating additional verification measures, account restrictions, and legal actions against manipulation services to ensure trustworthy consumer reviews.

\subsection{Emerging Trust Management Solutions in Specialised Review Crowdsourcing}
Recent literature proposes policies and algorithms to enhance trustworthiness in review crowdsourcing platforms by detecting misinformation and fraud.
A multi-layered model \cite{item125} incorporates user authentication, provider-platform cooperation, and automated content verification using service profiles. It integrates hybrid moderation (rule-based detection, human oversight), labels verified reviews, and sanctions fraud, leveraging supervised machine learning (e.g., hierarchical neural networks), unsupervised anomaly detection (e.g., One-Class Support Vector Machine (SVM)), and rule-based methods. Inputs—review texts, reviewer metadata, behaviour data—yield legitimate/fraudulent classifications with confidence scores.

A textual analysis approach \cite{item126} uses attribute salience, valence, and concreteness to assess authenticity. The supervised Indian Buffet Process (sIBP) and unsupervised topic modelling (via 'texteffect' R package) process reviews and trustworthiness scores to identify influential textual clusters, improving detection tools.

The DeepTrust framework \cite{item127} employs supervised Long Short-Term Memory Recurrent Neural Networks (LSTM-RNNs) to classify users as trustworthy, fraudulent, or unreliable, using temporal bipartite graphs and peer context. Inputs—reviews, timestamps, peer data—generate trustworthiness labels and behavioural embeddings, outperforming benchmarks (F1-score: 0.93, Area Under the Receiver Operating Characteristic curve (AUROC): 0.97).

The PriVs system \cite{item128} uses hybrid collaborative filtering (user-based and item-based) for app privacy recommendations, but this semi-supervised approach focuses on privacy, not misinformation or trust, limiting its relevance to review platforms.

These methods—multi-layered frameworks, textual analysis, and deep learning—collectively bolster trust and mitigate misinformation in review crowdsourcing.

\subsection{Insights from Specialised Review Crowdsourcing Trust Management}
\subsubsection{Strengths of Established Practices}
\begin{itemize} 
    \item \emph{Robust Automated and Community Validation.} 
    Platforms like Yelp, TripAdvisor, and Amazon combine proprietary machine learning algorithms with community-driven moderation, effectively detecting fraudulent or suspicious reviews. 
    
    \item \emph{Transparent User Profiles and Reputation Systems.} 
    Yelp’s ``Elite Squad'', TripAdvisor’s badges, and Amazon’s ``Verified Purchase'' labels clearly signal reviewer credibility, enhancing consumer trust and engagement. 
    
    \item \emph{Structured Reviewer Programs.} 
    Initiatives like Amazon Vine systematically generate trustworthy content by enforcing stringent guidelines and transparency, significantly improving review reliability. 
    
    \item \emph{Bidirectional Communication.} 
    Allowing verified business owners to publicly respond to reviews (Yelp, TripAdvisor) increases transparency, context-awareness, and accountability.
\end{itemize}

\subsubsection{Challenges and Limitations}
\begin{itemize} 
    \item \emph{Algorithm Transparency Issues.} 
    Proprietary algorithms that filter and rank reviews (e.g., Yelp, TripAdvisor) lack transparency, potentially leading to distrust or disputes over perceived fairness. 
    
    \item \emph{Vulnerability to Manipulation.} 
    Platforms remain susceptible to coordinated fraudulent activities such as review bombing or astroturfing, particularly due to limited visibility into algorithmic processes. 
    
    \item \emph{Limited Explicit Contributor Reputation Metrics.} 
    Many platforms lack clear, explicit public metrics indicating contributor reliability or historical accuracy, making it difficult for consumers to consistently discern trustworthy reviewers. 
    
    \item \emph{Contextual Sensitivity Limitations.} 
    Algorithms may inadequately account for regional differences or domain-specific contexts, potentially misclassifying authentic reviews as anomalous or vice versa.
\end{itemize}

\subsubsection{Opportunities for Future Advancement}
\begin{itemize} 
    \item \emph{Hybrid Verification Frameworks.} 
    Integrating multi-layered verification, combining supervised machine learning (e.g., hierarchical neural networks, LSTM-based temporal models) with rule-based detection, can enhance accuracy and fraud detection \cite{item125}, \cite{item127}. 
    
    \item \emph{Textual Feature-Based Authenticity Detection.} 
    Employing advanced computational text-mining techniques (e.g., supervised Indian Buffet Process for textual clustering) to systematically assess review authenticity can significantly enhance trust management \cite{item126}. 
    
    \item \emph{Adaptive Contributor Reputation Systems.} 
    Developing dynamic, transparent reputation metrics based on historical review accuracy, consistency, and community validation would incentivise high-quality contributions. 
    
    \item \emph{Enhanced Contextual Algorithms.} 
    Context-aware NLP and adaptive algorithms tuned to localised nuances can reduce false positives and improve the reliability of anomaly detection systems. 
    
    \item \emph{Increased Transparency in Moderation.} 
    Clarifying how reviews are flagged or filtered through publicly accessible dashboards or explanatory interfaces can reinforce user trust, engagement, and accountability.
\end{itemize}
By addressing these areas and leveraging existing strengths, specialised review crowdsourcing platforms can significantly enhance their trust management capabilities, ensuring sustained accuracy, reliability, and credibility in user-generated reviews.

\section{Overall Insights and Future Directions for Crowdsourcing Trust Management}

This study aimed to address three core research questions: identifying the strengths of current trust management practices on prominent crowdsourcing platforms, understanding the limitations and challenges these platforms face, and exploring opportunities for future advancement in trust management systems. Diverse crowdsourcing platforms, including volunteered geographic information, Wiki ecosystems, social media, mobile crowd-sensing initiatives, specialised environmental platforms, and review sites, exhibit distinct trust management frameworks aligned with their unique objectives, user bases, and data characteristics. Despite these variations, common themes, challenges, and avenues for improvement emerge from this analysis.

\subsection{Established Practices and Strengths}
\begin{itemize} 
    \item \emph{Community-Driven Validation.} 
    A pervasive strength across platforms is their reliance on the ``many eyes'' principle, where communal engagement acts as a primary mechanism for detecting, flagging, and correcting errors. Examples include OSM’s collaborative map editing, Wikipedia’s crowdsourced revisions, and Yelp’s community-driven review moderation. This user-driven scrutiny fosters a dynamic, self-correcting ecosystem. 
    
    \item \emph{Hybrid Moderation and Quality Checks.} 
    Platforms consistently employ automated and semi-automated moderation tools complemented by human oversight. Rule-based systems (e.g., Wikipedia's ClueBot NG), anomaly detection algorithms (Waze, OpenSignal), and machine learning-driven content filters (social media, review platforms) efficiently manage high volumes of data, while human moderators address nuanced issues. 
    
    \item \emph{Reputation and Role-Based Systems.} 
    Reputation elements, such as formal ranks (Waze), administrative privileges (Wikipedia), or elite badges (Yelp), incentivise quality contributions. Even informal reputation signals, like OSM’s area managers or Amazon's verified purchase tags, enhance trust by recognising proven reliability. 
    
    \item \emph{Transparency and Open Collaboration.} 
    Open data frameworks (e.g., OSM, Safecast) and transparent interaction opportunities (TripAdvisor’s business responses, MyCoast’s local authority verification) foster trust through accountability and collaborative validation. 
    
    \item \emph{Scalability Through Automation.} 
    Automated data collection and analysis enable platforms like OpenSignal and social media platforms (e.g., X) to manage vast datasets, providing rapid detection and response capabilities.
\end{itemize}

\subsection{Cross-cutting Challenges and Limitations}
\begin{itemize} 
    \item \emph{Scale Versus Depth.} 
    The balance between rapid automated detection and nuanced contextual validation remains difficult, especially with high volumes of daily contributions on platforms like social media and sensor networks, often overwhelming human moderators and automation alike. 
    
    \item \emph{Contributor Variability and Data Heterogeneity.} 
    Variability in user expertise, motivation, and data accuracy introduces inconsistency and bias. Platforms must manage issues ranging from ``review bombing'' on consumer review sites to sensor miscalibrations in environmental monitoring. 
    
    \item \emph{Opaque Algorithms and User Distrust.} 
    Proprietary filtering and ranking algorithms, while designed to prevent manipulation, may foster user distrust through a lack of transparency, potentially alienating legitimate contributors and reducing platform credibility. 
    
    \item \emph{Coverage Gaps and Uneven Engagement.} 
    Limited user engagement in niche or geographically isolated areas results in lower data quality due to insufficient oversight, creating challenges for platforms like TripAdvisor in under-visited locations or MyCoast in remote flood zones. 
    
    \item \emph{Power Concentration and Elite Dominance.} 
    Decision-making authority often concentrates among influential veteran contributors or moderators, risking systemic biases, reducing resilience, and undermining broad-based community trust. 
    
    \item \emph{Vulnerability to Sophisticated Attacks.} 
    Platforms remain susceptible to coordinated misinformation campaigns, targeted vandalism, and deliberate data manipulation, challenging existing trust management defences.
\end{itemize}
\subsection{Opportunities for Future Advancement}
\begin{itemize} 
    \item \emph{Adaptive and Transparent Reputation Frameworks.} 
    Developing dynamic, transparent reputation scores incorporating factors such as historical accuracy, peer endorsements, and contextual reliability can mitigate power concentration, equitably recognising diverse contributor efforts. 
    
    \item \emph{Context-Aware Machine Learning.} 
    Employing advanced AI methods (deep learning, transformers, graph neural networks) enhanced by external datasets (satellite imagery, official records) could significantly improve detection accuracy, capturing evolving misinformation patterns more effectively. 
    
    \item \emph{Cross-Verification and Data Fusion.} 
    Integrating crowdsourced data with authoritative references, sensor networks, or official records enhances validation, reduces erroneous submissions, and increases reliability. 
    
    \item \emph{Structured Community Feedback Loops.} 
    Extending structured feedback mechanisms like X’s Community Notes across various platforms encourages evidence-based corrections, decentralising authority and enhancing content quality through collective intelligence. 
    
    \item \emph{Transparent AI Explanation and Uncertainty Metrics.} 
    Providing explicit explanations and uncertainty measures alongside AI-driven trustworthiness assessments improves transparency, user comprehension, and trust.  
    
    \item \emph{Decentralised Moderation and Governance.} 
    Implementing decentralised moderation methods, such as rotating decision panels and community-elected oversight bodies, can diversify decision-making perspectives, reduce bias, and enhance systemic fairness and resilience. 
    
    \item \emph{Privacy-Preserving Trust Mechanisms.} 
    Leveraging cryptographic methods and blockchain technologies, particularly in mobile crowdsensing applications, can ensure data integrity and contributor privacy.
\end{itemize}

\subsection{A Soft Strategy for Continuous Trust Improvement}
An adaptive strategy combining advanced AI technologies and participatory governance offers a promising path forward for managing trust in crowdsourced data. Key components include:
\begin{itemize} 
    \item \emph{Real-Time Credibility Scoring.} 
    Immediate trustworthiness scores provided by AI, accompanied by clear explanations and uncertainty quantification. 
    
    \item \emph{Structured Community Engagement.} 
    Enabling community-driven validation, allowing user inputs to refine AI evaluations and enhance overall accuracy. 
    
    \item \emph{Transparent Dashboards.} 
    Publicly accessible dashboards displaying trust scores and decision rationales, fostering transparency and user comprehension. 
    
    \item \emph{Reinforcement Learning with Human Feedback.} 
    Continuous AI improvement guided by expert community feedback, ensuring robust decision-making without resorting to rigid moderation policies. 
    
    \item \emph{Decentralised Moderation.} 
    Rotating panels and modular, data-specific AI moderation systems that integrate diverse perspectives, mitigating biases and enhancing equity within platform governance.
\end{itemize}
This integrated approach ensures accountability, transparency, and sustained user engagement, preserving the collaborative ethos central to crowdsourcing platforms.

\subsection{Concluding Remarks}
Effective trust management is critical for harnessing crowdsourced data's full potential. By blending advanced technology with inclusive governance structures, platforms can address existing challenges, improve data quality, and foster sustained trust, ultimately translating collective intelligence into actionable, reliable insights across diverse applications.

\section{Acknowledgements}
This work was funded by DSTL, the UK Defence Science and Technology Laboratory, through the Defence Data Research Centre. We are grateful to Zena Wood, Glen Hart and Lucy Holford for their comments and feedback on this work.

\end{document}